\documentclass[journal]{IEEEtran}

\newcommand{\todo}[1]{\textcolor{red}{\textbf{TODO: #1}}}
\usepackage{xspace}
\newcommand{\toolname}{\textsc{CosmicTurtle}\xspace}
\usepackage{makecell}
\usepackage{numprint}
\usepackage{multirow}
\usepackage{booktabs}
\usepackage{comment}
\usepackage{siunitx}

\usepackage[table]{xcolor}
\usepackage[utf8]{inputenc}
\usepackage[colorlinks,urlcolor=blue,linkcolor=blue,citecolor=blue]{hyperref}
\usepackage{xspace}
\usepackage{ifthen}
\usepackage[T1]{fontenc}
\usepackage{amsmath,amssymb}
\usepackage{graphicx}
\usepackage{svg}
\usepackage{hyperref}
\usepackage{xcolor}
\usepackage{caption}
\usepackage{subcaption}
\usepackage{tabularx}
\usepackage{wrapfig}
\usepackage{balance}

\AtBeginDocument{%
  }

\begin{document}

\title{Verifiable Provenance of Software Artifacts with Zero-Knowledge Compilation                                                                                                                                                                                                                                                                                                                                                                                                                                                                                                                                                                                                                                                                                                                                                                                                                                                                                                                                                                                                                                                                                                                                                                                                                                                                                                                                                                                                                                                                                                                                                                                                                                                                                                          }



\author{
    \IEEEauthorblockN{Javier Ron, Martin Monperrus} \\    \textit{javierro@kth.se, monperrus@kth.se}\\
    \IEEEauthorblockA{KTH Royal Institute of Technology}
    
}

\maketitle

\begin{abstract}
  Verifying that a compiled binary originates from its claimed source code is a fundamental security requirement, called source code provenance. 
  Achieving verifiable source code provenance in practice remains challenging.
  The most popular technique, called reproducible builds, requires difficult matching and reexecution of build  toolchains and environments.
  We propose a novel approach to verifiable provenance based on compiling software with zero-knowledge virtual machines (zkVMs).
  By executing a compiler within a zkVM, our system produces both the compiled output and a cryptographic proof attesting that the compilation was performed on the claimed source code with the claimed compiler.
  We implement a proof-of-concept implementation using the RISC Zero zkVM and the ChibiCC C compiler, and evaluate it on 200 synthetic programs as well as 31 OpenSSL and 21 libsodium source files.
  Our results show that zk-compilation is applicable to real-world software and provides strong security guarantees: all adversarial tests targeting compiler substitution, source tampering, output manipulation, and replay attacks are successfully blocked.
\end{abstract}

\section{Introduction}

Establishing source code provenance, i.e., the ability to trace a binary back to the source files that were used to compile it and to demonstrate that the two correspond, is a fundamental security requirement for trustworthy software distribution~\cite{hassanshahi2025unlocking}.
Ensuring provenance of software artifacts is essential to mitigating supply chain attacks~\cite{ladisa_taxonomy, backstabbers} and maintaining public trust in open-source ecosystems.

Traditional compilation processes do not ensure provenance: when a user receives a compiled binary, there is no guarantee that they are also given the source code and that it was actually produced from this claimed source code.
It is even uncommon to know the exact compiler that was used to compile the binary~\cite{thompson1984reflections}.
This creates significant security risks, as malicious actors could distribute binaries that differ from their purported source, potentially containing backdoors or other malicious code.

More recently, the reproducible build concept~\cite{9403390} has gained popularity, defined as independent parties being able to regenerate identical binaries from the same source, with byte-for-byte equality as strong evidence of provenance.
Yet, reproducibility is brittle and expensive to maintain, especially for complex toolchains or non-deterministic build environments~\cite{canonicalization, randrianaina2024options}.
Complementary efforts explore trusted execution environments (TEEs) to attest compilation and build processes~\cite{delignat2023should, Hugenroth2025AttestableBC}, but these approaches require specialized hardware and shift trust to hardware vendors.
In this paper, the problem we are addressing is to provide strong guarantees of software provenance without the hurdle of full rebuilding and without relying on trusted proprietary hardware.

\begin{figure}[t]
    \centering
    \includegraphics[width=0.95\linewidth]{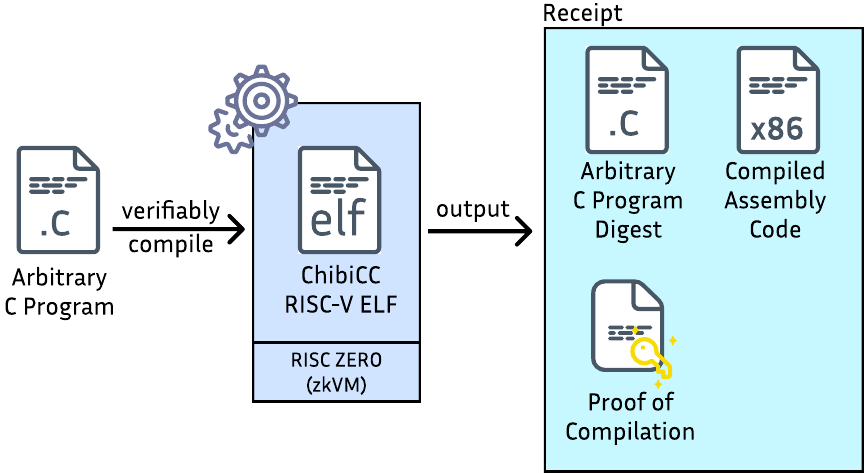}
    \caption{Provenance guarantees in \toolname. The compiler is executed inside RISC Zero's zkVM. It records the computation trace while compiling the source program. It then generates a succinct cryptographic proof attesting that the compilation occurred on the claimed source code with the claimed compiler, i.e. cryptographically verifiable provenance.}
    \label{fig:instrumented_execution}
\end{figure}

We propose a novel approach to provenance based on verifiable compilation through zero-knowledge virtual machines (zkVMs).
By executing a compiler within a zkVM, our system produces both the compiled output --the binary-- and a cryptographic proof attesting that the compilation ran on the claimed source code with the claimed compiler.
Our approach provides cryptographic guarantees of provenance that can be independently verified by anyone without reproducing the build or trusting specialized hardware.
Because zkVMs execute standard binaries, the approach is generally applicable to any compiler without requiring compiler modifications.

We evaluate our design on a mix of real-world and synthetic programs: 31 libsodium source files (crypto library), 21 OpenSSL source files (crypto library), as well as 200 randomly generated C programs produced by Csmith \cite{csmith}.
All 252 programs are successfully zk-compiled and verified with cryptographic proofs of provenance.
Our systematic adversarial tests confirm that the core security properties of provenance hold: compilers cannot be substituted, source code cannot be tampered with, binary code also, and replay attacks are all detected and rejected by the verification process.
In other words, our experiments fully validate our concept of zero-knowledge compilation. 

The main contributions of this work are:
\begin{itemize}
    \item A novel approach for software provenance based on verifiable compilation with zero-knowledge virtual machines. Our approach produces a cryptographic proof of compilation, binding source code, compiler identity, and compiled output.
    \item A threat model and security analysis identifying the assumptions, threats, and guarantees of our approach.
    \item A proof-of-concept implementation using the RISC Zero zk virtual machine and the ChibiCC C compiler, demonstrating practical feasibility, publicly available at \url{https://github.com/ASSERT-KTH/verifiable-compilation}.
    \item An empirical evaluation using 200 synthetic C programs, 21 OpenSSL as well as 31 libsodium source code files, clearly demonstrating the applicability and security guarantees.
\end{itemize}

\section{Background on Verifiable Execution}

Verifiable execution refers to mechanisms that allow a party to convince others that a computation was performed correctly without requiring them to re-run the entire computation.
The fundamental challenge of verifiable execution is balancing the cost of verification against the strength of the guarantee: stronger assurance typically requires more computational effort.
In the context of compilation, verifiable execution would enable a prover to demonstrate that a specific compiler transformed a given source program into a particular binary, thereby establishing the provenance of the resulting artifact.
There are three main lines of research in verifiable execution.

\subsubsection{Re-execution}
\textit{Re-execution} is the most direct approach to verifying computation: a verifier simply re-runs the program on the claimed inputs and checks that the outputs match.
For compilation, this translates to independently rebuilding a binary from its source with an identical toolchain, also known as reproducible builds ~\cite{9403390}.
However, the approach makes strong assumptions: (1) access and execution of the full build toolchain, (2) deterministic build environments, and (3) computational resources comparable to the original build.
These prerequisites are often not met in practice.
Any deviation such as non-deterministic compilers, undisclosed patches, or environment skew yield mismatches~\cite{docker,canonicalization}.

\begin{figure*}[t]
    \centering
    \includegraphics[width=0.99\textwidth]{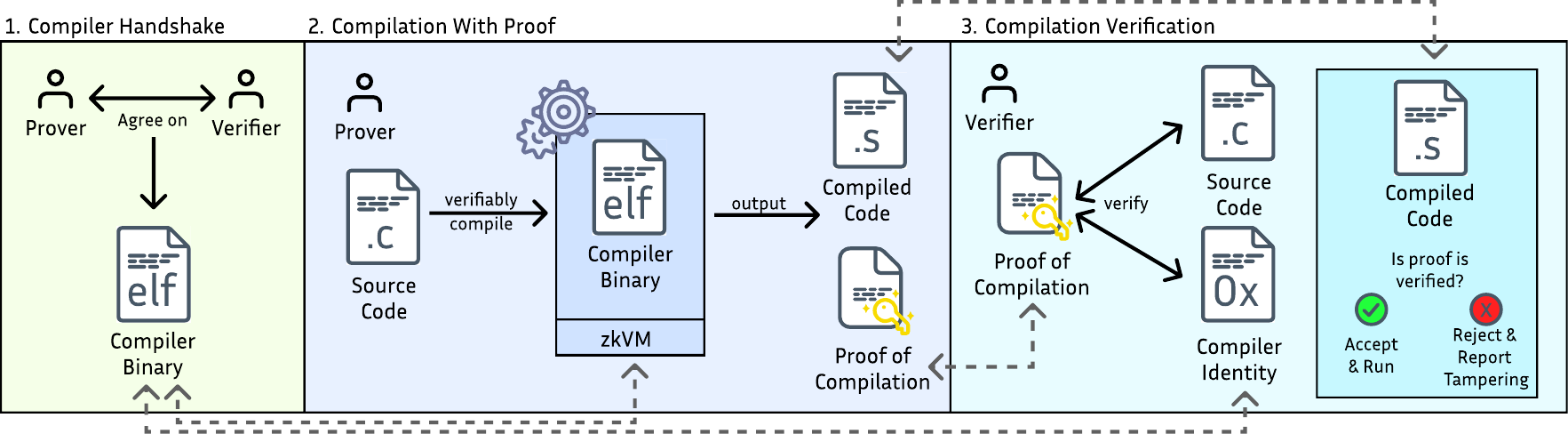}

    \caption{System overview showing the three phases of verifiable compilation: (1) compiler handshake where prover and verifier agree on the compiler binary and its cryptographic identity, (2) compilation with proof where the prover executes the compiler within a zkVM to generate both the binary output and a cryptographic proof, and (3) verification where the verifier validates the proof against the committed software artifacts, i.e. validates that the binary is correctly obtained from the claimed source code and compiler. Once verified, the verifier has integrity guarantees to run the program.}
    \label{fig:overview}
  \end{figure*}

\subsubsection{Trusted Execution Environments}
\textit{Trusted execution environments} (TEEs) amortize verification by delegating trust to hardware enclaves that attest to the software they execute.
TEEs provide isolated execution environments protected by hardware mechanisms, ensuring that code running within an enclave cannot be tampered with or observed by the host operating system or other processes~\cite{tee}.
Major TEE implementations include Intel Software Guard Extensions (SGX)~\cite{sgx} and AMD Secure Encrypted Virtualization (SEV)~\cite{amd_sev}, each providing different isolation guarantees and attestation mechanisms.

The core security property of TEEs is \textit{remote attestation}: the ability to produce a cryptographically signed statement, an \textit{attestation quote}, that proves a specific piece of software has run in a genuine TEE instance.
This attestation typically includes measurements of the code, e.g., cryptographic hashes; the TEE's identity; and optionally the inputs and outputs of the computation~\cite{tee_meng}.
In a verifiable compilation setting, the compiler runs inside an enclave, which emits the binary plus an attestation quote binding the source and toolchain measurements~\cite{Hugenroth2025AttestableBC}.
Verifiers can then validate the attestation quote against the TEE vendor's certificate authority, confirming that the claimed compilation occurred within a genuine enclave.
TEEs avoid re-running the computation, but they shift trust to hardware vendors and require enclave-compatible toolchains.

\subsubsection{Zero-Knowledge Virtual Machines}
A zkVM executes programs compiled to a specific instruction set which has zero-knowledge execution semantics~\cite{cryptoeprint:2023/1032}, producing a cryptographic proof that the execution was performed correctly.
A verifier can check this proof in time shorter than re-running the computation, without access to the program's internal state.

Internally, a zkVM \textit{arithmetizes} each execution trace: it encodes every instruction, register update, and memory access as a system of polynomial constraints.
A valid proof demonstrates that some assignment to these constraints exists and is consistent with the claimed public inputs and outputs.

ZkVMs proofs have two important properties:
\textit{Succinctness} means the proof is short and verification is cheap ~\cite{oude2024systematic}.
\textit{Zero-knowledge} means the proof reveals nothing beyond the validity of the statement; this property is optional and can be enabled when intermediate state must remain private.
In this paper, our primary interest is succinctness: we need a compact witness that the compilation occurred, that is used later to verify provenance.

zkVMs have found applications in settings where a prover must convince resource-limited verifiers that a large computation executed correctly: in blockchain rollups~\cite{fAmulet,chaliasos2024analyzing} and in delegated cloud computation~\cite{chiesa2015cluster}. To our knowledge, we are the first to study the use of zero-knowledge execution in the context of compilation and software provenance.

\section{Threat Model}
\label{sec:threats}

Our system is organized around two abstract roles: (1) a \textit{prover} that produces compilation artifacts and (2) a \textit{verifier} that validates them.
They interact over four main artifacts: 
(1) a source program,
(2) a compiler (binary executable).
(3) the compiled assembly code, 
and (4) a proof of compilation.
We now define a threat model that captures provenance-relevant threats based on those two roles and four artifacts.

\subsection{Threats}
\begin{figure}
    \centering
    \includegraphics[width=0.95\linewidth]{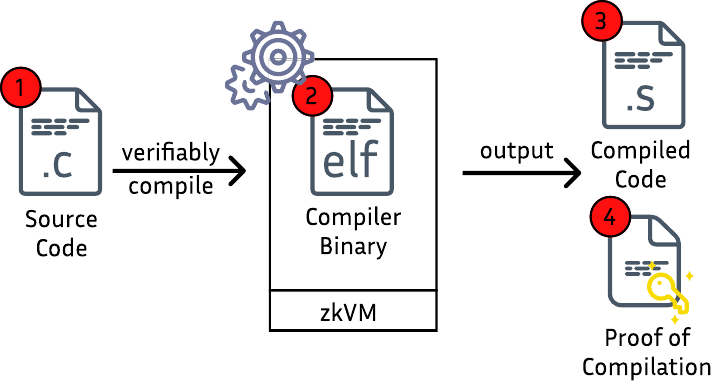 }
    \caption{Threat model for verifiable compilation. The prover generates compilation artifacts while the verifier validates them. Four threats target the main artifacts: (1) source code tampering before or during compilation, (2) compiler substitution with a malicious binary, (3) binary code tampering after compilation, and (4) replay attacks reusing valid proofs with different artifacts.}
    \label{fig:threat-model}
\end{figure}

Our system aims to protect against the following specific threats, each corresponding to an artifact, as shown in~\autoref{fig:threat-model}:

1) \textit{Source Code Tampering}: An adversary modifies the source code before or during compilation, producing a binary that differs from the claimed source~\cite{przymus2025wolves}.
This threat encompasses several attack vectors: modification of source files in transit between the developer and the build system, tampering with source code repositories, or man-in-the-middle attacks during source code retrieval.
The adversary's goal is to introduce malicious functionality while maintaining the appearance that the binary corresponds to benign, publicly auditable source code.
Unlike compiler substitution, source code tampering targets a specific compilation rather than compromising the toolchain itself.

2) \textit{Compiler Tampering}: An adversary replaces the legitimate compiler binary after agreement between prover and verifier with a malicious one that produces backdoored or incorrect output.
This threat is particularly insidious because the malicious compiler could produce functionally correct output for most inputs while injecting subtle vulnerabilities or backdoors in specific cases~\cite{thompson1984reflections}.
The adversary may control the build infrastructure, supply chain, or distribution channel through which the compiler is obtained.
The impact of successful compiler substitution is severe: all binaries produced by the malicious compiler could be compromised, potentially affecting downstream users who believe they are using legitimate software.

3) \textit{Binary Code Tampering}: An adversary modifies the compiled assembly output after compilation but before distribution to end users.
This threat targets the distribution phase, where an attacker with access to build artifacts could replace legitimate binaries with malicious versions.
The adversary may compromise build servers, artifact repositories, or content delivery networks.
Output manipulation is particularly dangerous when combined with legitimate-looking metadata, as users may trust the binary based on its apparent provenance.

4) \textit{Replay Attacks}: An adversary reuses a valid proof from a previous compilation with a different or modified output.
In this scenario, the adversary possesses a legitimately generated proof for one compilation but attempts to associate it with a different binary.
This could occur when distributing updated (potentially malicious) versions of software while presenting proofs generated for earlier, legitimate versions.
The adversary exploits the temporal gap between proof generation and verification, hoping that verifiers will accept stale proofs for fresh artifacts.

\subsection{Assumptions}
Our threat model relies on the following foundational assumptions:

First, \textit{Proof System Soundness}: We assume the underlying zero-knowledge proof system is cryptographically sound.
Formally, this means that for any probabilistic polynomial-time adversary, the probability of producing a valid proof for a false statement is negligible in the security parameter.
This assumption relies on well-studied cryptographic hardness assumptions, such as the discrete logarithm problem or collision-resistant hash functions, depending on the specific proof system employed.

Second, \textit{zkVM Correctness}: We assume the zkVM implementation correctly arithmetizes the instruction set semantics into constraint systems.
The zkVM must faithfully capture all state transitions, register updates, memory accesses, and control flow decisions during execution, ensuring that the proof corresponds to an actual execution trace.
This assumption encompasses both the correctness of the constraint generation (the zkVM produces constraints that accurately encode the instruction semantics) and the completeness of the encoding (no execution behaviors are omitted from the constraint system).
Bugs in the zkVM implementation could allow adversaries to produce proofs for executions that did not actually occur or that violated the expected semantics.

Third, \textit{Prover and Verifier Coordination}: We assume that the public artifacts, including the zkVM implementation and verification systems, are agreed upon by both prover and verifier and used consistently before proof generation and verification.
This assumption requires an out-of-band mechanism for establishing consensus on these artifacts, such as publication through trusted channels, cryptographic signing by known authorities.

Violations of these assumptions are outside the scope of this work.

\section{System Design}

Given the threat model of \autoref{sec:threats}, we propose a novel design for verifiable compilation that establishes provenance through zero-knowledge virtual machines.

\subsection{Verifiable Compilation Workflow}

The execution flow operates in three phases: compiler handshake, compilation with proof, and compilation verification as depicted in \autoref{fig:overview}.

1) \textit{Compiler Handshake.} First, the prover and the verifier need to agree on the compiler binary to use.
This is can be done in two ways: (1) both the prover and verifier take the same compiler source code and produce the compiler binary, both resulting binaries should be bit-for-bit identical; or (2) the prover produces the compiler binary and shares it with the verifier.
In both cases, the prover and the verifier agree on the compiler binary and its cryptographic identity.
The second option is practical and avoids the infinite recursion loop ~\cite{verifiable_compilation}. 

2) \textit{Compilation with Proof.} During the compilation with proof phase, the prover executes the compiler within a zkVM that records the entire computation trace.
The compiler receives a source program as input and produces assembly code as output, while the zkVM constructs a constraint system encoding the execution.
The zkVM's constraint system captures the relationship between the input source code, the compiler's identity, and output assembly code, along with the intermediate state transitions that occurred during compilation.
The zkVM then generates a succinct cryptographic proof attesting that the assembly output was indeed produced by running the claimed compiler on the claimed input source code, i.e. a proof of compilation.

3) \textit{Verification.} During verification, the verifier receives these artifacts and validates the proof using the zkVM's verifier algorithm, which confirms that the proof corresponds to a valid execution of the compiler agreed at handshake.
Successful verification establishes provenance: the assembly code was compiled from the declared source using the declared compiler without requiring the verifier to reproduce the build (to rerun compilation).
After successful verifation, the verifier can proceed with running the program. 
Failed verification means that one step or artifact has been tampered with, causing the verifier to reject and report the binary program as potentially compromised.
The compromised compiled program is never executed and thus does not cause any damage. 

\subsection{Usage In Practice}
In practice, the prover and verifier roles map to different software industry stakeholders depending on the deployment context.

In an \textit{open-source distribution} scenario, the prover is an open source build service (e.g. a Linux distribution's build farm) that compiles source code and publishes binaries along with compilation proofs; and verifiers are end users (e.g. a medical tech company running Linux) or third party security auditors who validate that distributed binaries match their claimed source repositories.

In a \textit{software vendor} scenario, the prover is the software company that builds and ships proprietary applications; verifiers are enterprise customers or regulators who require cryptographic assurance that the deployed binary corresponds to audited source code.

\subsection{Security Properties}
Our verifiable compilation system achieves four security properties, each directly countering a threat from \autoref{sec:threats}:

1) \textit{Source Code Integrity} counters \textit{Source Code Tampering} (Threat 1):
The proof commits to a cryptographic digest of the source code as a public input to the zkVM execution.
Any modification to the source, whether in transit, at rest, or during compilation, produces a different digest, causing verification to fail.
Verifiers independently hash the claimed source and compare it against the committed digest in the proof.

2) \textit{Compiler Integrity} counters \textit{Compiler Tampering} (Threat 2):
The proof is bound to the compiler's cryptographic identity, which is derived from the compiler binary's content.
Substituting a different compiler, even one producing identical outputs for most inputs, changes the identity, causing the verifier to reject the proof.
This binding is established during the handshake phase and enforced cryptographically during verification.

3) \textit{Output Integrity} counters \textit{Binary Code Tampering} (Threat 3):
The compiled output is committed within the proof as a public output of the zkVM execution.
Any post-compilation modification to the binary breaks the cryptographic binding between the proof and the output, causing verification to fail.

4) \textit{Proof Freshness} counters \textit{Replay Attacks} (Threat 4):
The soundness of the underlying proof system ensures that valid proofs can only be generated for actual executions with specific inputs and outputs.
Since proofs are cryptographically bound to both the source code and the compiled output, reusing a proof from a previous compilation with a different source code or output fails verification.
The one-to-one binding between proof, source code, compiler, and output prevents any mix-and-match attacks.

\section{Prototype Implementation}
This section explains the concrete tooling choices and implementation details of our prototype verifiable compilation system, which we call \toolname.

\subsection{Core Components}

1) \textit{zkVM}. We use RISC Zero as zero-knowledge virtual machine\footnote{\url{https://www.risczero.com/}}. RISC Zero is a zero-knowledge virtual machine that executes RISC-V programs and generates succinct proofs of execution.
RISC Zero's architecture makes it particularly well-suited for our verifiable compilation approach, as it can execute arbitrary RISC-V programs (such as a compiler) while maintaining the ability to generate efficient proofs.
The RISC-V instruction set provides a clean, standardized target that simplifies the compilation process and enables interoperability with existing toolchains.

2) \textit{Compiler}. We use
ChibiCC\footnote{\url{https://github.com/rui314/chibicc}}, a C compiler developed as an educational project, with x86 as target.
ChibiCC is well-suited for our proof-of-concept implementation for three reasons: (1) it is self-contained with minimal external dependencies, making it reasonable to cross-compile to RISC-V; (2) its clean, readable codebase of approximately 10,000 lines of C facilitates debugging and adaptation for zkVM execution; and (3) despite its simplicity, it supports a substantial subset of the C language sufficient for compiling real-world programs (see \autoref{sec:real-world}).
We use commit \texttt{d367510f} of ChibiCC that supports core C features including structs, unions, pointers, arrays, control flow, and function calls.

\subsection{Implementation Details}
\label{sec:implementation}
We now describe the implementation of each phase of the verifiable compilation workflow.

1) \textit{Compiler Handshake.}
To build ChibiCC for execution within RISC Zero, we cross-compile it to the RISC-V instruction set, specifically targeting the \texttt{riscv32im} architecture (32-bit RISC-V with integer multiplication extension).
We use the official RISC-V toolchain's \texttt{riscv32-unknown-elf-gcc} compiler to produce a standalone ELF binary.
The resulting binary is the compiler artifact that will execute inside the zkVM.

To compute the compiler's cryptographic identity, we use RISC Zero's \texttt{risc0\_zkvm::compute\_image\_id} function, which computes a hash of the ELF binary's code and data segments; this hash is called the \textit{ImageID}.
The ImageID uniquely identifies the compiler binary and is used during verification to ensure the correct compiler was executed.
This is the one that the prover and the verifier agree upon.

2) \textit{Compilation with Proof.}
The prover executes the ChibiCC RISC-V binary within RISC Zero's zkVM, as illustrated in \autoref{fig:instrumented_execution}.
The source program is provided as input to the zkVM through its standard input mechanism, and the compiler writes the generated assembly to the zkVM's output buffer.
During execution, the zkVM records every instruction, register state, and memory access into an execution trace.

Upon completion, the zkVM produces a cryptographic artifact called a \textit{receipt}.
The receipt contains three components: (1) the cryptographic proof attesting to correct execution, (2) the SHA-256 digest of the input source code, and (3) the compiled assembly output.
The proof cryptographically binds all contents of the receipt to a valid execution trace, ensuring that the claimed outputs were actually produced by running the committed compiler on the committed inputs.
The receipt is serialized and can be distributed alongside the compiled code as a portable provenance attestation.

3) \textit{Verification.}
The verification process  uses RISC Zero's verification logic to validate the receipt against the agreed-upon compiler identity.
Verification requires inputs: 
the source code, 
the receipt,
and the compiler ImageID established during the handshake phase.

The verifier first deserializes the receipt and extracts the committed input digest and output assembly.
The verifier then computes the SHA-256 hash of the claimed source code and compares it against the input digest in the receipt; a mismatch indicates that the source code was tampered with or differs from what was compiled.
If the digests match, the verifier invokes RISC Zero's \texttt{receipt.verify(image\_id)} function, which cryptographically validates that the proof corresponds to a valid execution of the binary identified by the ImageID.
Successful verification confirms that the assembly output was produced by executing the agreed-upon compiler on the claimed source code.
Failed verification causes the verifier to reject the assembly output and report it as potentially compromised.


\section{Experimental Methodology}
To validate the provenance guarantees of our verifiable compilation system, we employ two techniques: (1) compiling random programs, and (2) compiling real-world open-source program.

\subsection{Random program Generation}
\textit{Random program generation} provides a systematic approach to testing compiler correctness by exploring diverse program structures, control flows, and language features that might not be adequately covered by hand-written test suites~\cite{donaldson2025solidity}.
We use Csmith~\cite{csmith}, a well-established tool for compiler testing that generates random, well-formed C programs conforming to the C standard. Csmith is an ideal choice for evaluating our system's ability to handle a wide variety of compilation scenarios.
Csmith allows us to assess that the zkVM-based proof generation remains sound and efficient even when processing programs with varying sizes, control structures, and data types.

To evaluate our system, we use Csmith to generate a corpus of 200 random C programs.
Each generated program is compiled using our verifiable compilation pipeline \toolname, which executes ChibiCC within the RISC Zero zkVM environment.
We generated 200 random C programs using Csmith, and compiled them with \toolname within the RISC Zero zkVM environment.
For each program, we measured the compilation and proof generation time, verification time, source file size, and resulting proof (receipt) size.
We also compile the generated programs using a standard run of the ChibiCC compiler to compare the compilation time.

We configure Csmith to generate programs with restricted complexity to ensure compatibility with the subset of C supported by our ChibiCC version.
Specifically, we disable features such as volatiles, packed structs, and bitfields that are not supported by the target version of ChibiCC.

\subsection{Real-World Programs}
\label{sec:real-world}
To demonstrate that our approach generalizes beyond synthetic benchmarks, we also compile source files extracted from real-world open-source projects.
We select individual C source files from two widely-used cryptographic libraries: libsodium\footnote{\url{https://github.com/jedisct1/libsodium}} and OpenSSL\footnote{\url{https://github.com/openssl/openssl}}.
These projects represent security-critical domains where verifiable compilation would provide significant value.
We manually select files that are compatible with our version of ChibiCC, avoiding files that rely on unsupported language features such as inline assembly, complex preprocessor macros, or GNU extensions.
Our goal is to successfully compile at least 10 programs from these real-world libraries.

\textbf{Security Validation.}
To validate the security properties defined in \autoref{sec:threats}, we conduct adversarial tests simulating each threat scenario:
\begin{itemize}
    \item \textit{Compiler substitution}: We replace the compiler binary with a modified version (different ImageID) and attempt verification with the original ImageID. Verification should fail.
    \item \textit{Source code tampering}: We modify the source code after the handshake phase and attempt to verify using the original source digest. Verification should fail due to digest mismatch.
    \item \textit{Output manipulation}: We modify the compiled assembly output after proof generation and attempt verification. Verification should fail due to broken cryptographic binding.
    \item \textit{Replay attacks}: We attempt to reuse a valid proof from one compilation with a different source file or output. Verification should fail due to mismatched commitments.
\end{itemize}

\subsection{Setup}
Our experimental setup runs on an Intel i9-10980XE CPU, with 128GB of RAM, and 2x NVIDIA GeForce RTX 2080 Ti GPUs. We use RISC Zero version 3.0.0 at commit \texttt{a84bb194} and ChibiCC commit \texttt{d367510f}.

\section{Experimental Results}

\subsection{Compilation with Proof}
We generated 200 random C programs using Csmith, and compiled them with \toolname within the RISC Zero zkVM environment.
We also applied \toolname to a subset of C files from libsodium and OpenSSL, both being crypto libraries underpinning the Internet with high integrity requirements.

For each program, we measured the source file size, standard compilation time, zk-compilation time (compilation with proof), verification time, and receipt size.
Table~\ref{tab:summary_stats} summarizes these metrics across the three corpora.
For random programs, source sizes range from 0.14~KB to 83~KB, with zk-compilation times between \numprint{57.01}~s and \numprint{979.12}~s and a median of \numprint{78.91}~s.
The OpenSSL and libsodium files exhibit similar behavior: despite differing code bases, their zk-compilation times remain in the same order of magnitude, with medians of \numprint{80.55}~s and \numprint{82.28}~s, respectively (Table~\ref{tab:summary_stats}).
Across all three datasets, receipt sizes grow proportionally with source size, from hundreds of kilobytes for the smallest programs to over 100~MB for the largest, indicating that storage costs scale with program complexity.

The zk-compilation overhead compared to standard compilation is substantial.
For random programs, the median zk-compilation time (\numprint{78.91}s) is roughly four orders of magnitude slower than the median standard compilation time (\numprint{0.0092}s) reported in Table~\ref{tab:summary_stats}.
This overhead reflects the current state of zkVM technology: the zkVM must arithmetize every instruction into polynomial constraints and generate a cryptographic proof over the entire execution trace.
However, this cost is paid once per build and amortized over all subsequent verifications; moreover, zk-proving performance has historically improved quickly and is expected to continue to do so~\cite{chaliasos2024analyzing}.

\begin{table*}[h]
\centering
\begin{tabular}{l  | S[table-format=2.4] S[table-format=2.4] S[table-format=4.4] | S[table-format=2.4] S[table-format=2.4] S[table-format=4.4] | S[table-format=2.4] S[table-format=2.4] S[table-format=4.4]}
\toprule
\multicolumn{1}{c}{\multirow{2}{*}{\textbf{Metric}}} & \multicolumn{3}{c}{\textbf{Random Programs (200)}} & \multicolumn{3}{c}{\textbf{OpenSSL (31)}} & \multicolumn{3}{c}{\textbf{libsodium (21)}} \\
\cmidrule(lr){2-4} \cmidrule(lr){5-7} \cmidrule(lr){8-10}
{} & {Min} & {Median} & {Max} & {Min} & {Median} & {Max} & {Min} & {Median} & {Max} \\
\midrule
C File Size (KB)          & 0.14 & 2.37 &  82.94 & 0.42 & 1.85 & 8.57 & 0.19 & 1.21 & 19.91 \\
Receipt Size (MB)         & 0.26 & 2.79 & 114.41 & 0.21 & 0.52 & 2.74 & 0.25 & 0.52 & 9.15  \\
\midrule
Standard Compile Time (s) & 0.0006 & 0.0092 & 0.0665  & 0.0012   & 0.0021   & 0.0046    & 0.0007 & 0.0023 & 0.0124    \\
zk-Compile Time (s)       & 57.01  & 78.91  & 979.12  & 59.71 & 80.55 & 104.83 & 68.25 & 82.28 & 137.21 \\
\midrule
Verify Time (s)           & 47.93  & 49.88  & 68.75   & 48.12 & 48.64 & 50.64  & 48.17 & 48.81 & 51.34  \\

\bottomrule
\end{tabular}
\caption{Summary statistics across 200 randomly generated test programs, 31 OpenSSL files, and 21 libsodium files.}
\label{tab:summary_stats}
\end{table*}

Figure~\ref{fig:combined_results} presents our experimental results across key dimensions.
Subfigure (a) shows the relationship between program size and compilation time.
The results demonstrate a linear relationship between source code size and compilation time.
The trend line indicates predictable compilation durations with \toolname.

Subfigure (b) illustrates how the size of generated proofs correlate with the source program size.
The smallest proofs (approximately 0.27 MB) were generated for trivial programs, while the largest (120 MB) corresponded to the most complex test case with 83 KB of source code. The disk requirements for proof storage are also predictably linear.

Finally, we note that the cryptographic routines from libsodium and OpenSSL, involving arithmetic operations and control flow, and found that they did not introduce any anomalies in proof generation or verification.
This demonstrates that compilation with \toolname is applicable to real-world code in a security-critical domain.

\subsection{Provenance Verification}

Table~\ref{tab:summary_stats} also reports verification times for all programs.
For random Csmith programs, verification takes between \numprint{47.93}~s and \numprint{68.75}~s, with a median of \numprint{49.88}~s, while OpenSSL and libsodium files show similarly tight ranges with medians of \numprint{48.64}~s and \numprint{48.81}~s, respectively.
Unlike zk-compilation time, verification time varies only mildly with program size: once the proof is generated, checking it is a small fraction of the original computation length.

\begin{figure*}[t]
    \centering
    \includegraphics[width=\textwidth]{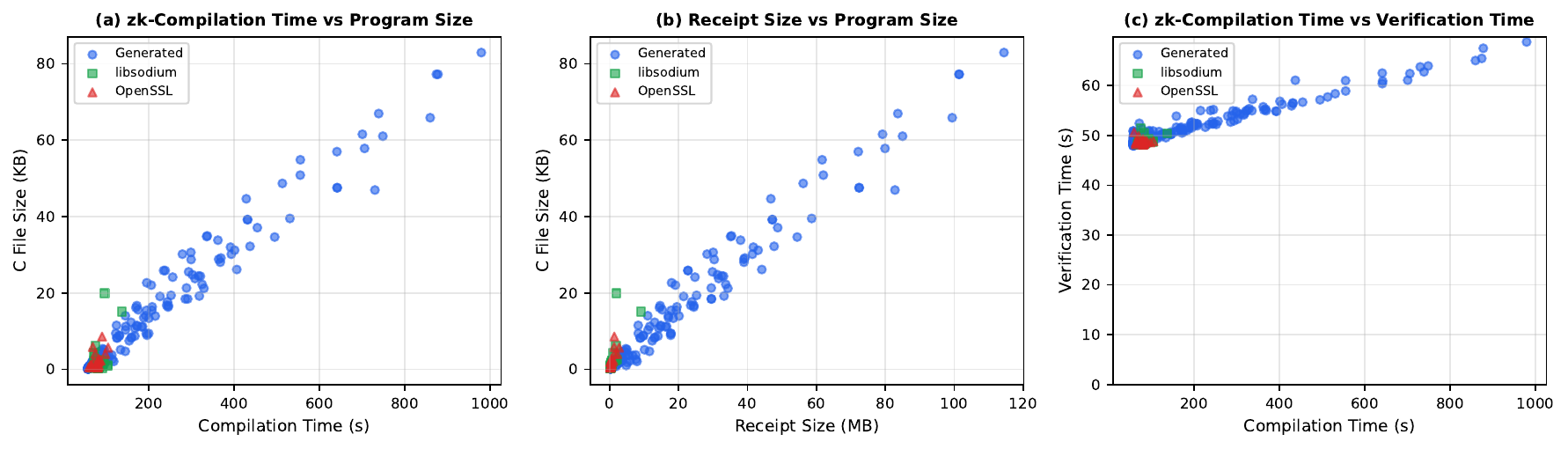}
    \caption{Experimental results across randomly generated programs from Csmith and real-world programs from libsodium and OpenSSL (C file sizes in KB, receipt sizes in MB). (a) Compilation time vs program size shows linear scaling. (b) Receipt size vs program size demonstrates reasonable storage requirements for the generated proofs.  (c) Compilation time vs verification time shows verification is consistently faster than proof generation.}
    \label{fig:combined_results}
\end{figure*}

Figure~\ref{fig:combined_results}, subfigure (c), directly compares zk-compilation time against verification time.
Across all datasets, verification is consistently faster than zk-compilation, confirming that the succinctness properties of the proof system are realized in practice.
This separation between expensive one-time proof generation and comparatively cheap repeated verification is precisely what makes zkVM-based verifiable compilation attractive for distribution scenarios with many verifiers but few provers.

\subsection{Security Validation}

We validated the security properties of our system by conducting adversarial tests on the libsodium and OpenSSL programs.
For each program, we attempted to subvert the verification process according to the threat scenarios defined in \autoref{sec:threats}.

\textbf{Compiler Substitution.}
We modified the ChibiCC binary by making a minimal change to the compiler code, which caused it to produce a new ImageID.
When attempting to verify receipts generated with the original compiler against the modified ImageID, verification failed in all cases.
This confirms that the cryptographic binding between proof and compiler identity prevents compiler substitution attacks.

\textbf{Source Code Tampering.}
We modified the source code of each program by changing a function name before proof generation, then recomputed the source digest.
In all cases, verification failed because the recomputed digest did not match the digest committed in the receipt.
This confirms that any modification to the source code after proof generation is detectable.

\textbf{Output Manipulation.}
We modified the compiled assembly output by changing a function name after proof generation.
In all cases, verification failed because the modified output did not match the output committed in the receipt.
This confirms that post-compilation tampering is detectable.

\textbf{Replay Attacks.}
We attempted to reuse a valid receipt from one compilation with a different source file.
In all cases, verification failed because the source digest in the receipt did not match the digest of the claimed source.
This confirms that proofs cannot be replayed across different compilations.

Overall, all tests confirmed that the security properties hold as expected.

\subsection{Summary of Experimental Findings}

Our evaluation demonstrates that zkVM-based verifiable compilation is a feasible approach to establishing provenance for real-world C programs.
The key findings are:

\textbf{Performance.}
The zk-compilation overhead is substantial: median zk-compilation time is four orders of magnitude slower than median standard compilation time.
This overhead stems from the zkVM's need to arithmetize every instruction into polynomial constraints and generate a cryptographic proof over the entire execution trace. This is a one-time overhead.

\textbf{Applicability.} The system successfully generated and verified proofs for all 200 randomly generated test programs, 31 programs in OpenSSL and 21 programs in libsodium, demonstrating robustness across diverse program structures.

\textbf{Security.} All four security properties held across all adversarial tests: compiler substitution, source code tampering, output manipulation, and replay attacks were detected and rejected by the verification process, validating the threat model guarantees.

These results validate that our approach provides a practical foundation for verifiable compilation, offering strong cryptographic guarantees of provenance.

\section{Discussion}

\subsection{Regulatory Compliance}
High-stakes software domains, e.g., medical devices, aerospace systems, financial infrastructure, and defense applications, face stringent regulatory requirements for software provenance and traceability.
Regulations such as the FDA's guidance on software validation~\cite{fda_software_validation}, DO-178C~\cite{do_178c} for airborne systems, and financial sector requirements like PCI-DSS~\cite{pci_dss} mandate documented evidence that deployed software corresponds to reviewed and approved source code, along with auditable build and release processes.

Our verifiable compilation approach provides a technical mechanism that directly supports these obligations.
Proofs can be generated once during the official build process and archived alongside other compliance artifacts, then verified repeatedly by auditors, regulators, or customers without access to the build infrastructure.
The succinct nature of verification makes it feasible to integrate provenance checks into continuous compliance pipelines.
The cryptographic binding between source, compiler, and output  provides the precise traceability regulators seek: given a deployed binary, stakeholders can verify which source code and toolchain produced it.

\subsection{On Zero-Knowledge}
Although our approach leverages zero-knowledge virtual machines, it is important to clarify that the \textit{zero-knowledge} property itself is not essential to verifiable compilation.
What our system fundamentally requires is a \textit{succinct proof of correct execution}: a compact cryptographic witness demonstrating that a specific computation (compilation) was performed correctly on claimed inputs (source code, compiler) to produce claimed outputs (assembly or binary).
In an open-source or high-integrity contractual context, all parties have access to the three core elements (the source code,
 the compiler, the binary output) thus, zero-knowledge is not required. 
In other words, all inputs and outputs are public and intentionally revealed to the verifier; there are no secrets to hide.
For us, the zkVM serves as a general-purpose proof generator.


\section{Related Work}

\textit{Binary-source matching}:
Several approaches attempt to recover provenance links between binaries and source code through static analysis.
For instance, BinPro traces function-level similarities between binaries and source repositories to recover provenance links after compilation has stripped away high-level structure~\cite{Miyani2017BinProAT}.
In the same spirit, BinaryAI is a similar approach that uses machine learning to match binaries to source code repositories~\cite{jiang2024binaryai}.
These techniques provide post-hoc analysis but cannot guarantee that a binary was produced from specific source code; they only establish similarity or likelihood.
Our approach differs fundamentally by providing a cryptographic proof generated during compilation, rather than inferring speculative provenance after the fact.

\textit{Compilation on TEEs}:
Hugenroth et al. \cite{Hugenroth2025AttestableBC} propose a system for compiling verifiable binaries using trusted execution environments. In contrast to our work, their solution depends on specialized TEE hardware and trust in TEE vendors.
Our work instead is hardware independent, and vendor independent.
We reduce the trust assumptions to the soundness of the underlying proof system and the correctness of its implementation, both being full open source.

\textit{Formal verification}:
Formal verification is a technique for verifying the correctness of a system by proving that it satisfies a set of properties.
CompCert~\cite{leroy2016compcert} is a prominent example of a formally verified C compiler proving that the compiler correctly translates source code to assembly.
While CompCert guarantees that the specified properties hold, it does not guarantee provenance.

\textit{Proof-carrying code}:
Proof-carrying code (PCC), introduced by Necula and Lee~\cite{necula1996safe,necula1997pcc}, addresses a complementary problem: ensuring that executable code satisfies safety properties, e.g., type safety, memory safety, without trusting the code producer.
In PCC, compiled code is accompanied by a machine-checkable proof that the code adheres to a specified safety policy; the consumer verifies this proof before execution, establishing trust in the code's behavior.
Our approach differs in its goal: rather than proving properties about what the code does, we prove facts about how the code was produced.
PCC establishes behavioral safety guarantees, while verifiable compilation establishes provenance guarantees.
The two techniques are complementary: a certifying compiler~\cite{necula1998design} could run inside a zkVM, producing both a safety proof and a compilation proof, thereby attesting both that the binary is safe and that it originated from specific source code.

\section{Conclusion}
We have presented a novel approach to software provenance based on verifiable compilation using zero-knowledge virtual machines.
By executing compilers within zkVMs, we can generate cryptographic proofs that attest to the provenance of a given binary.
Our approach provides strong security guarantees for software provenance related to non-tampering of code, compilers and binaries.
Our approach addresses critical trust issues in software supply chain distribution and assessment.
Our experimental evaluation demonstrates the feasibility of verifiable compilation using existing zkVM technology.
We believe that zero-knowledge execution and verification can also be useful for solving fundamental integrity problems 
in software packaging, package registries and 
vulnerability disclosure.

\balance
\bibliographystyle{unsrt}
\bibliography{citations}

\end{document}